\documentclass[prl,aps,showpacs,byrevtex,twocolumn,groupedaddress]{revtex4}
\usepackage{amsfonts}
\usepackage{graphicx}
\usepackage{amsmath}

\input{tcilatex}
\begin{document}

\title{Appearance of Gauge Fields and Forces beyond the adiabatic
approximation}
\author{Pierre Gosselin$^{1}$ and Herv\'{e} Mohrbach$^{2}$}
\affiliation{$^{1}$Institut Fourier, UMR 5582 CNRS-UJF, UFR de Math\'{e}matiques,
Universit\'{e} Grenoble I, BP74, 38402 Saint Martin d'H\`{e}res, Cedex,
France }
\affiliation{$^{2}$Laboratoire de Physique Mol\'{e}culaire et des Collisions, ICPMB-FR
CNRS 2843, Universit\'{e} Paul Verlaine-Metz, 57078 Metz Cedex 3, France}

\begin{abstract}
We investigate the origin of quantum geometric phases, gauge fields and
forces beyond the adiabatic regime. In particular, we extend the notions of
geometric magnetic and electric forces discovered in studies of the
Born-Oppenheimer approximation to arbitrary quantum systems described by
matrix valued quantum Hamiltonians. The results are illustrated by several
physical relevant examples.
\end{abstract}

\maketitle

A physical system can never be considered as completely isolated from the
rest of the universe. For a slow (adiabatic) cyclic variation of its
environment, the wave function of the quantum system gets an additional
geometric phase factor, known as the Berry phase \cite{BERRY1}. In fact, the
driving environment, the 'heavy' or 'slow' system, is also subject to back
reaction from the 'light' or 'fast' system. In the context of the
Born-Oppenheimer theory of molecules, the back reaction of the light system
leads to the appearance of a gauge field in the effective Hamiltonian for
the slow one (the environment) \cite{MEAD}\cite{MOODY}\cite{ZYGELMAN}. The
gauge field consists of a vector and a scalar potential and turns out to
depend on a quantum geometric tensor \cite{BERRY2}. It can both induce
interference phenomena and modify the dynamics through geometric Lorentz and
electric forces \cite{BERRY3}\cite{STERN}.

In this note we investigate the origin of quantum gauge fields and forces in
a more general context, by considering the diagonalization of an arbitrary
matrix valued quantum Hamiltonian. To be precise, by diagonalization it is
meant the derivation of an effective in-band Hamiltonian made of
block-diagonal energy subspaces. For that purpose we use the results of a
powerful method developed recently \cite{PIERRE1}. This approach based on a
new differential calculus on non-commutative space, where $\hbar $ plays the
role of running parameter, leads to an in-band energy operator that can be
obtained systematically up to arbitrary order in $\hbar .$ Particularly
important for our purpose, it is possible to give an explicit effective
arbitrary diagonal Hamiltonian to order $\hbar ^{2}$ in terms of
non-canonical coordinates and commutators between gauge fields (Eq. $\left(
79\right) $ in \cite{PIERRE1}). We will directly apply this result to
systems whose Hamiltonian has the simple form $H=T(\mathbf{K})+V(\mathbf{Q})$
and the components are assumed to fulfill the canonical commutation
relations $\left[ Q_{i},K_{j}\right] =i\hbar \delta _{ij}$. We then discuss
how gauge fields arise in physical situations as various as Dirac and Bloch
electrons in electric fields or Born-Oppenheimer theory. Note that there
exists another totally different method of diagonalization in a formal
series expansion in $\hbar $ which uses symbols of operators via Weyl
calculus \cite{LITTLEJOHN}. To our knowledge this method was only applied to
a Born-Oppenheimer-type Hamiltonian \cite{LITTLEJOHN}.

Our approach reveals the appearance at order $\hbar ^{2}$ of a scalar gauge
potential expressed in terms of two tensors. One is the quantum metric
tensor \cite{BERRY2}\cite{PROVOST}, and the other is a new tensor
generalizing an additional term found in \cite{LITTLEJOHN} for the
Born-Oppenheimer case. Another very important consequence of the Hamiltonian
diagonalization is the appearance of gauge invariant intraband coordinates.
The advantage of using these coordinates is that the diagonal Hamiltonian is
also gauge invariant. Moreover, these coordinates fulfill a non-commutative
algebra which strongly affects the dynamics through a Lorentz term and the
gradient of a new scalar potential, generalizing thus the adiabatic dynamics
of the Born-Oppenheimer theory.

\textit{Hamiltonian diagonalization. }Consider a Hamiltonian of the form 
\begin{equation}
H=T(\mathbf{K})+V(\mathbf{Q})  \label{HTV}
\end{equation}%
where we assume that $H$ has a matrix representation with $T(K)$
non-diagonal and $V(Q)$ diagonal. In \cite{PIERRE1}, by considering $\hbar $
as a running parameter, we relate the in-band Hamiltonian $%
UHU^{+}=\varepsilon \left( \mathbf{X}\right) $ and the unitary transforming
matrix $U\left( \mathbf{X}\right) $ (where $\mathbf{X\equiv (\mathbf{Q},%
\mathbf{K})}$) to their classical expressions through integro-differential
operators, i.e. $\varepsilon \left( \mathbf{X}\right) =\widehat{O}\left(
\varepsilon _{0}\left( \mathbf{X}_{0}\right) \right) $ and $U\left( \mathbf{X%
}\right) =\widehat{N}\left( U_{0}\left( \mathbf{X}_{0}\right) \right) $,
where matrices with the subscribe $0$ correspond to operators replaced by
classical commuting variables $\mathbf{X}_{0}\mathbf{=}\left( \mathbf{Q}_{0}%
\mathbf{,K}_{0}\right) .$ The only requirement of the method is therefore
the knowledge of $U_{0}\left( \mathbf{X}_{0}\right) $ which gives the
diagonal form $\varepsilon _{0}\left( \mathbf{X}_{0}\right) .$ Generally,
these equations do not allow to find directly $\varepsilon \left( \mathbf{X}%
\right) $, $U\left( \mathbf{X}\right) $, however, they allow us to produce
the solution recursively in a series expansion in $\hbar .$ With this
assumption that both $\varepsilon $ and $U$ can be expanded in power series
of $\hbar ,$ we determined in Eq. $\left( 79\right) $ of \cite{PIERRE1}, the
explicit diagonalization of an arbitrary Hamiltonian to order $\hbar ^{2}$.
The expression of the effective $n$-th in-band energy $\varepsilon _{n}$
greatly simplifies for Hamiltonian given by Eq. $\left( \ref{HTV}\right) .$
Indeed, the first order is easily obtain by an unitary transformation $U_{0}(%
\mathbf{K})$ diagonalizing $T(\mathbf{K})$ giving $U_{0}HU_{0}^{+}=%
\varepsilon _{0}(\mathbf{K})+V(\mathbf{Q+\hbar A})$ with $\mathbf{A}\equiv
iU_{0}\nabla _{\mathbf{K}}U_{0}^{+}.$ Then $V$ is now non-diagonal. The
diagonalization at the next order is done by an unitary transformation
matrix $U=U_{0}+\hbar U_{0}U_{1}$ with the antihermitian matrix $%
(U_{1})_{mn}=\frac{\left( 1-\delta _{mn}\right) }{\varepsilon _{m}\left(
t\right) -\varepsilon _{n}\left( t\right) }(\mathbf{A)}_{mn}.\nabla V,$
which removes the off-diagonal elements of $\mathbf{A}$ and leads to
corrections of order $\hbar ^{2}$ such that $\varepsilon _{n}$ in Eq. $%
\left( 79\right) $ of \cite{PIERRE1} becomes 
\begin{equation}
\varepsilon _{n}\left( \mathbf{K,q}_{n}\right) =\varepsilon _{0,n}\left( 
\mathbf{K}\right) +V\left( \mathbf{q}_{n}\right) +\hbar ^{2}\Phi _{n}
\label{epsilonh2}
\end{equation}%
where the geometric scalar potential is 
\begin{equation}
\Phi _{n}(\mathbf{\mathbf{Q,K)}}=\frac{G_{n}^{ij}}{2}\mathbf{\partial }_{i}%
\mathbf{\partial }_{j}V(\mathbf{\mathbf{Q})+}T_{n}^{ij}\mathbf{\partial }%
_{i}V(\mathbf{\mathbf{Q})}\partial _{j}V(\mathbf{Q})  \label{scalar}
\end{equation}%
with two gauge invariant tensors $G_{n}^{ij}$ and $T_{n}^{ij}$ defined as 
\begin{equation}
G_{n}^{ij}(\mathbf{K})=\frac{1}{2}\sum\nolimits_{m\neq n}\left(
(A^{i})_{nm}(A^{j})_{mn}+h.c.\right)  \label{metric}
\end{equation}%
and%
\begin{equation}
T_{n}^{ij}(\mathbf{\mathbf{K}})=\frac{1}{2}\sum\nolimits_{m\neq n}(\frac{%
(A^{i})_{nm}(A^{j})_{mn}}{\varepsilon _{0,m}-\varepsilon _{0,n}}+h.c.)
\label{T}
\end{equation}%
The tensor $G_{n}^{ij}$ is known as the quantum metric tensor \cite{BERRY2}%
\cite{PROVOST} and $T_{n}^{ij}$ is a new tensor generalizing an additional
term found in \cite{LITTLEJOHN} for the Born-Oppenheimer theory. In Eq. $%
\left( \ref{epsilonh2}\right) $ we introduced the intraband coordinate%
\textbf{\ }$\mathbf{q}_{n}=\mathbf{Q}+\hbar \mathbf{a}_{n}$\textbf{\ }where%
\textbf{\ }$\mathbf{a}_{n}$\textbf{\ }is a gauge connection usually called
the Berry connection defined as\textbf{\ }$\mathbf{a}_{n}(\mathbf{K})\equiv (%
\mathbf{A})_{nn}=i\left\langle n\right. \left\vert \nabla _{\mathbf{K}%
}n\right\rangle $.\textbf{\ }Here\textbf{\ }$\left\vert n\right\rangle $%
\textbf{\ }are the eigenstates of the non-diagonal part of\textbf{\ }$H,$%
\textbf{\ }i.e.,\textbf{\ }$T(\mathbf{K})\left\vert n\right\rangle
=\varepsilon _{0,n}(\mathbf{K})\left\vert n\right\rangle $\textbf{. }The
introduction of the non-canonical coordinate\textbf{\ }$\mathbf{q}_{n}$%
\textbf{\ }essential to maintain the gauge invariance of the Hamiltonian,
implies non-canonical commutation relations $\left[ q_{n}^{i},q_{n}^{j}%
\right] =i\hbar ^{2}\Theta _{n}^{ij}$ with\textbf{\ }$\Theta _{n}^{ij}(%
\mathbf{K})=\frac{\partial a_{n}^{j}}{\partial K_{i}}-\frac{\partial
a_{n}^{i}}{\partial K_{j}}+\left[ a_{n}^{i},a_{n}^{j}\right] $\textbf{\ }the
Berry gauge curvature in the $n$-th eigenstate. The Heisenberg equations of
motion to the second order in $\hbar $ are 
\begin{equation}
\overset{\cdot }{\mathbf{q}}_{n}=\nabla _{\mathbf{K}}\varepsilon _{n}-\frac{%
\hbar }{2}(\overset{\cdot }{\mathbf{K}}\times \mathbf{\Theta }_{n}-\mathbf{%
\Theta }_{n}\times \overset{\cdot }{\mathbf{K}}),\text{ \ }\overset{\cdot }{%
\mathbf{K}}=-\nabla _{\mathbf{q}_{n}}V  \label{EQMVT}
\end{equation}%
where we introduced the "magnetic field" $\Theta _{n}^{i}=\varepsilon
_{jk}^{i}\Theta _{n}^{jk}$. The dynamics of the intraband operators leads
directly to a Lorentz-type term. The scalar potential is a consequence of
transitions between eigenstates and impacts the dynamics through its
gradient. Working with the non-canonical coordinates is a short-cut to
determine the dynamics of a system prepared in an eigenstate of the full
Hamiltonian. This state will evolve in the same energy subspace as far as we
can neglect higher contributions in the expansion in $\hbar $. In
comparison, the equations of motion derived from the Hamiltonian $H$ do not
seem to include a Lorentz force, and the determination of the
"eigendynamics" can be a very difficult to achieve. An appealing example is
given in \cite{BERRY3}\cite{STERN} where the "exact" slow motion of a
massive neutral particle coupled to a spin is compared with the
Born-Oppenheimer theory.\textbf{\ }

We underline that our diagonalization does not need the adiabatic
assumption, because it is an "exact" diagonalization. However, the expansion
in $\hbar $ breaks down in regions of mode conversion where $\varepsilon
_{0,m}-\varepsilon _{0,n}<<\hbar $ or for large values of\textbf{\ }$%
(A^{i})_{nm}=i\left\langle m\right. \left\vert \partial _{i}n\right\rangle .$%
\textbf{\ }Now, if in a particular regime the probability transition between
different eigenstates is very small, on can neglect with a good
approximation the off-diagonal elements.\textbf{\ }This is usually
considered as an adiabatic approximation and we see that it coincides with
the semiclassical approximation. In a mode conversion region, one can easily
generalize the diagonalization of $H$\ to a block-diagonalization allowing
transitions between eigenstates inside the block. In this case, we can
therefore consider the semiclassical limit without having adiabaticity.%
\textbf{\ }

\textit{Born-Oppenheimer approximation. }Consider the following Hamiltonian
describing a fast system in interaction with an external environment 
\begin{equation}
H=\frac{1}{2}B_{ij}P^{i}P^{j}+\frac{p^{2}}{2m}+\varphi (\mathbf{R,r})
\end{equation}%
where the fast system is described by a set of dynamical variables $(\mathbf{%
r,p})$ and the slow one by coordinates $(\mathbf{R,P}).$ As in \cite{BERRY2}
we consider a general kinetic energy with $B$, a positive definite inverse
mass tensor. Applying the previous results with the mapping $\mathbf{Q}%
\rightarrow \mathbf{P}$, $\mathbf{K}\rightarrow \mathbf{R}$ (and $i\nabla _{%
\mathbf{K}}\rightarrow -i\nabla _{\mathbf{R}}$) we have $V(\mathbf{Q}%
)\rightarrow B_{ij}P_{i}P_{j}/2$ and we obtain the following eigenvalues for
the slow system (assuming a non-degenerate spectrum for the fast system)

\begin{equation}
\varepsilon _{n}=\frac{1}{2}B_{ij}p_{n}^{i}p_{n}^{j}+\hbar ^{2}\Phi
_{n}+E_{n}(\mathbf{R})
\end{equation}%
where $\mathbf{p}_{n}=\mathbf{P}-i\hbar \left\langle n\right. \left\vert
\nabla _{\mathbf{R}}n\right\rangle $ and $\left\vert n(\mathbf{R}%
)\right\rangle $ is the eigenstate of the fast Hamiltonian with energy $%
E_{n}(\mathbf{R})$. The scalar potential Eq. $\left( \ref{scalar}\right) $
then becomes 
\begin{equation}
\Phi _{n}(\mathbf{R,P})=\frac{G_{n}^{ij}(\mathbf{R})}{2}B_{ij}\mathbf{+}%
T_{n}^{ij}(\mathbf{R})B_{il}B_{jk}P^{l}P^{k}
\end{equation}%
with the quantum metric tensor $G_{n}^{ij}(\mathbf{R})=\func{Re}\sum_{m\neq
n}\left\langle \partial _{i}n\right. \left\vert m\right\rangle \left\langle
m\right. \left\vert \partial _{j}n\right\rangle $ and $T_{n}^{ij}(\mathbf{R}%
)=\func{Re}\sum_{m\neq n}\frac{\left\langle \partial _{i}n\right. \left\vert
m\right\rangle \left\langle m\right. \left\vert \partial _{j}n\right\rangle 
}{\varepsilon _{0,m}-\varepsilon _{0,n}}.$ The term $G_{n}^{ij}(\mathbf{R}%
)B_{ij}$ is the usual part of the scalar potential discussed in several
circumstances \cite{BERRY2}\cite{BERRY3}\cite{STERN}, whereas the term $%
T_{n}^{ij}(\mathbf{R})B_{il}B_{jk}P^{l}P^{k}$ was found in \cite{LITTLEJOHN}%
. Here we see that the Born-Oppenheimer theory can be obtained
straightforwardly from our Hamiltonian diagonalization to order $\hbar ^{2}.$
In the same manner from Eq.$\left( \ref{EQMVT}\right) $\ we immediately get
the Born-Oppenheimer equations of motion $\overset{\cdot }{\mathbf{p}}%
_{n}=-\nabla _{\mathbf{R}}E_{n}-\frac{\hbar }{2}(\overset{\cdot }{\mathbf{R}}%
\times \mathbf{\Theta }_{n}-\mathbf{\Theta }_{n}\times \overset{\cdot }{%
\mathbf{R}})-\hbar ^{2}\nabla _{\mathbf{R}}\Phi _{n}$ with $\overset{\cdot }{%
R}_{i}=B_{ij}p_{n}^{j}$. Similar equations of motion for a classical system
consisting of a classical magnetic moment interacting with an inhomogeneous
magnetic field \cite{BERRY3}\cite{STERN} were studied in details. It was
found that the Lorentz force results from a slight misalignment of the
magnetic moment relative to the magnetic field. This corresponds to the
semi-classical approximation. The electric force is a time average of a
strong oscillatory force induced by the precession of the magnetic moment.
This is a kind of zitterbewegung effect.

\textit{Particle in a linear potential. }Another interesting relevant
situation concerns a particle in a linear potential exemplified here by a
Bloch electron in an constant external electric field (see also ref. \cite%
{HANSSEN}). Consider $H=H_{0}(\mathbf{P,R})+\varphi (\mathbf{R})$ with $%
H_{0} $ the energy of a particle in a periodic potential and $\varphi (%
\mathbf{R})=-e\mathbf{E.R}$ the external electric perturbation (and $e<0$
the charge). Using the mapping $\mathbf{Q}\rightarrow \mathbf{R,}$ the
scalar gauge potential reduces to $\Phi _{n}(\mathbf{k}%
)=e^{2}T_{n}^{ij}E_{i}E_{j}$, and the energy eigenvalues are 
\begin{equation}
\varepsilon _{n}=\varepsilon _{0,n}\left( \mathbf{k}\right) -e\mathbf{%
E.r_{n}+}e^{2}\hbar ^{2}T_{n}^{ij}\left( \mathbf{k}\right) E_{i}E_{j}
\label{TIJ}
\end{equation}%
with $\varepsilon _{0,n}\left( \mathbf{k}\right) $ is the $n$-th energy band
and $\mathbf{k}$ the pseudo-momentum. The intraband position operator is $%
\mathbf{r}_{n}=\mathbf{R}+i\hbar \left\langle u_{n}\right. \left\vert \nabla
_{\mathbf{k}}u_{n}\right\rangle $ with $\left\vert u_{n}(\mathbf{k}%
)\right\rangle $ the periodic part of the Bloch wave function and $%
T_{n}^{ij}(\mathbf{k})=\func{Re}\sum_{m\neq n}\frac{\left\langle \partial
_{i}u_{n}\right. \left\vert u_{m}\right\rangle \left\langle u_{m}\right.
\left\vert \partial _{j}u_{n}\right\rangle }{\varepsilon _{0,m}-\varepsilon
_{0,n}}$. Introducing the "magnetic field" $\mathbf{\omega }_{n}=\frac{\hbar 
}{eE^{2}}\mathbf{E\times }\nabla _{\mathbf{k}}\Phi _{n}$ and $\mathbf{\chi }%
_{n}=\frac{1}{eE^{2}}\mathbf{E.}\nabla _{\mathbf{k}}\Phi _{n}$ the equations
of motion are%
\begin{equation*}
\overset{\cdot }{\mathbf{r}}_{n}=\nabla _{\mathbf{k}}\varepsilon _{0,n}-%
\frac{\hbar }{2}(\overset{\cdot }{\mathbf{k}}\times \Omega _{n}+\Omega
_{n}\times \overset{\cdot }{\mathbf{k}})+\hbar ^{2}\mathbf{\chi }_{n}\overset%
{\cdot }{\mathbf{k}},\ \ \overset{\cdot }{\mathbf{k}}=e\mathbf{E}
\end{equation*}%
where $\Omega _{n}=\mathbf{\Theta }_{n}+\hbar \mathbf{\omega }_{n}.$ This
shows that $\Phi _{n}$ contributes to the Lorentz term $\hbar \overset{\cdot 
}{\mathbf{k}}\times \Omega _{n}$ known as the anomalous velocity which is
orthogonal to the applied electric field. This anomalous velocity is at the
center of many\ recent experimental and theoretical works. $\Phi _{n}$
contributes also to the velocity in the direction of $\mathbf{E,}$ through
the term $\hbar ^{2}\mathbf{\chi }_{n}\overset{\cdot }{\mathbf{k}}.$

\textit{Berry phase. }The linear potential case has another interest. It
allows us to also consider the fast system and derive the Berry phase in a
different way. Indeed, consider a time dependent Hamiltonian $H(t)$ and
introduce the differential operator $D=H\left( t\right) -P_{0}$ where $%
P_{0}\equiv i\hbar \partial /\partial t$ is the conjugate of time which is
treated formally as an operator such that $\left[ P_{0},t\right] =i\hbar $.
The time dependence is due to the time evolution of some parameters $x(t)$\
describing the environment. To transform the system of differential
equations (Schr\"{o}dinger equation) $D\left\vert \Psi (t)\right\rangle =0$,
which couples all components of $\left\vert \Psi (t)\right\rangle $ into a
decoupled set of differential equations, we introduce an unitary
transformation $\left\vert \Psi ^{\prime }\left( t\right) \right\rangle
=U(t)\left\vert \Psi \left( t\right) \right\rangle $ such that $U(t)D\left(
t,P_{0}\right) U^{+}(t)=\widetilde{\Lambda }(t,P_{0})$ is a diagonal
differential operator and $\widetilde{\Lambda }(t,P_{0})\left\vert \Psi
^{\prime }\left( t\right) \right\rangle =0$. Therefore the time evolution is
given by $\left\vert \Psi ^{\prime }\left( t\right) \right\rangle =e^{\frac{%
-i}{\hbar }\int\nolimits_{0}^{t}\Lambda (t)dt}\left\vert \Psi ^{\prime
}\left( 0\right) \right\rangle .$ Since $\Lambda (t)=\widetilde{\Lambda }%
(t)+P_{0}$ is diagonal, no time ordered product is required. Returning back
to the initial state we have%
\begin{equation}
\left\vert \Psi \left( t\right) \right\rangle =U^{+}(t)e^{\frac{-i}{\hbar }%
\int\nolimits_{0}^{t}\Lambda (t)dt}U(0)\left\vert \Psi \left( 0\right)
\right\rangle   \label{PHI}
\end{equation}%
A system prepared in a state $\left\vert \Lambda _{n}(0)\right\rangle $
which is an eigenstate of $D$, i.e., $D(0)\left\vert \Lambda
_{n}(0)\right\rangle =\widetilde{\Lambda }_{n}(0)\left\vert \Lambda
_{n}(0)\right\rangle $ will evolve with $\Lambda (t)$ and thus stays in the
instantaneous eigenstates $\left\vert \Lambda _{n}(t)\right\rangle $ of $D(t)
$ (for simplicity we assume non degenerate eigenvalues). In this case the
wave function becomes $\left\vert \Psi \left( t\right) \right\rangle =e^{%
\frac{-i}{\hbar }\int\nolimits_{0}^{t}\Lambda _{n}(t)dt}\left\vert \Lambda
_{n}(t)\right\rangle $. Since eigenstates of $D$ instead of $H$ are
considered, the time evolution Eq. $\left( \ref{PHI}\right) $ is
non-adiabatic. In general we need an approximation scheme for the
diagonalization of $D$ and we will use the expansion to order $\hbar ^{2}.$
The problem of finding $\Lambda _{n}=\widetilde{\Lambda }_{n}(t)+P_{0}$ is
formally equivalent to the Bloch electron example discussed above with $%
K\rightarrow t,$ $Q\equiv R\rightarrow $ $P_{0}$ and $eE=1$. We obtain from
Eq.$\left( \ref{TIJ}\right) $

\begin{equation}
\Lambda _{n}=\varepsilon _{n}\left( t\right) -i\hbar \left\langle
n\left\vert \overset{\cdot }{n}\right\rangle \right. \mathbf{+}\hbar
^{2}T_{n}\left( t\right)   \label{Dn}
\end{equation}%
where $\varepsilon _{n}\left( t\right) $ and $n(t)$ are instantaneous
eigenvalues and eigenstates of $H$, i.e., $H(t)\left\vert n(t)\right\rangle
=\varepsilon _{n}\left( t\right) \left\vert n(t)\right\rangle $, and $%
T_{n}\left( t\right) =\func{Re}\sum_{m\neq n}\frac{\left\langle \overset{%
\cdot }{n}\right. \left\vert m\right\rangle \left\langle m\right. \left\vert 
\overset{\cdot }{n}\right\rangle }{\varepsilon _{m}-\varepsilon _{n}}.$
Therefore for a periodic motion of period $T$, not necessarily adiabatic,
such that $\left\vert \Lambda _{n}(T)\right\rangle =\left\vert \Lambda
_{n}(0)\right\rangle $ (single valued eigenstates), we have 
\begin{equation}
\left\vert \Psi \left( T\right) \right\rangle =e^{-i\gamma _{n}}\left\vert
\Psi \left( 0\right) \right\rangle   \label{phiT}
\end{equation}%
\textbf{with}%
\begin{equation*}
\gamma _{n}=\frac{1}{\hbar }\int\nolimits_{0}^{T}\varepsilon
_{n}(t)dt+i\int_{0}^{T}\left\langle n\left\vert \overset{\cdot }{n}%
\right\rangle dt\right. \mathbf{+}\hbar \int_{0}^{T}T_{n}\left( t\right) dt
\end{equation*}%
The phase $\gamma _{n}$\ appears as an expansion in power of $\hbar $. The
first term is the usual dynamical phase and the second one the geometric
Berry phase independent of $\hbar $\ and of the velocity of parameters $%
\overset{\cdot }{x}(t)$. The additional phase $\hbar \int_{0}^{T}T_{n}\left(
t\right) dt$\ of order $\hbar $\ is non-geometric as it depends on $\overset{%
\cdot }{x}$. It cancels in the infinitely slow $\overset{\cdot }{x}%
\rightarrow 0$\ adiabatic regime, which thus coincides with the
semiclassical approximation.\textbf{\ }Diagonalization at order $\hbar ^{2}$
thus goes beyond the adiabatic approximation and takes into account
transitions between eigenstates. Quantitatively, if the system is prepared
in an eigenstate $\left\vert n(0)\right\rangle $ of $H(0)$, then $\left\vert
\Psi (t)\right\rangle $ is given by Eq. $\left( \ref{PHI}\right) $ with $%
U=U_{0}+\hbar U_{1}U_{0}$ where $U_{1}\left( t\right) _{mn}=i\frac{\left(
1-\delta _{mn}\right) \left\langle m\left\vert \overset{\cdot }{n}%
\right\rangle \right. }{\varepsilon _{m}\left( t\right) -\varepsilon
_{n}\left( t\right) }$, so that we have the following expansion up to order $%
\hbar $:%
\begin{eqnarray*}
\left\vert \Psi (t)\right\rangle  &=&e^{\frac{-i}{\hbar }\int%
\nolimits_{0}^{t}\Lambda _{n}(t)dt}\left\vert n(t)\right\rangle +\hbar
\sum_{m\neq n}\left( e^{\frac{-i}{\hbar }\int\nolimits_{0}^{t}\Lambda
_{n}(t)dt}A_{mn}(t)\right.  \\
&&-\left. e^{\frac{-i}{\hbar }\int\nolimits_{0}^{t}\Lambda
_{m}(t)dt}A_{m_{n}}(0)\right) \left\vert m(t)\right\rangle +O(\hbar ^{2})
\end{eqnarray*}%
The magnitude of transitions is then controlled by the term $A_{mn}=\frac{%
i\left\langle m(t)\left\vert \overset{\cdot }{n}(t)\right\rangle \right. }{%
\varepsilon _{n}\left( t\right) -\varepsilon _{m}\left( t\right) }$ which is
neglected in the adiabatic limit $\left\langle m(t)\left\vert \overset{\cdot 
}{n}(t)\right\rangle \right. \rightarrow 0.$ Note that $\left\vert \Psi
(t)\right\rangle $ is normalized to unity at order $\hbar $ only. A
normalization at a higher order needs an expansion of $U$ to the same order 
\cite{PIERRE1}.

In principle deviation from adiabaticity could be measured by
interferometry. Consider a periodic two states system, and write the initial
state in the eigenbase $\left\vert n(0)\right\rangle =\left\vert \Lambda
_{n}(0)\right\rangle +\hbar A_{nm}(0)\left\vert \Lambda _{m}(0)\right\rangle
.$ Then, after one cycle $\left\vert \Psi \left( T\right) \right\rangle
=e^{-i\gamma _{n}}\left\vert \Lambda _{n}(0)\right\rangle +\hbar e^{-i\gamma
_{m}}A_{nm}(T)\left\vert \Lambda _{m}(0)\right\rangle .$ For an observable $%
O $ which does not commute with $H$ one will find in the average $%
\left\langle \Psi \left( T\right) \right\vert O\left\vert \Psi \left(
T\right) \right\rangle $ an interference term $2\hbar \func{Re}\left(
A_{nm}(T)\left\langle \Lambda _{n}(0)\right\vert O\left\vert \Lambda
_{m}(0)\right\rangle e^{-i\left( \gamma _{n}-\gamma _{m}\right) }\right) $
which would signal deviation from adiabaticity. As discussed below, this
interference effect is formally equivalent to the zitterbewegung of Dirac
particles.

\textit{Dirac particle in an external potential. }

We will now show that our formalism can also be used for relativistic Dirac
particles, which are usually treated with the Foldy Wouthuysen approach \cite%
{FOLDY}. The Hamiltonian is (with $c=1$)%
\begin{equation}
H=\mathbf{\alpha .p}+\beta m+V\left( \mathbf{R}\right)
\end{equation}%
where $\mathbf{\alpha }$ and $\beta $ are the usual $\left( 4\times 4\right) 
$ Dirac matrices and $V\left( \mathbf{R}\right) $ is the external potential.
The matrix diagonalizing the free part of the Hamiltonian $U_{0}\left( 
\mathbf{\alpha .p}+\beta m\right) U_{0}^{+}=\beta E$ with $E=\sqrt{\mathbf{p}%
^{2}+m^{2}}$ is the usual Foldy Wouthuysen unitary transformation $U_{0}=%
\frac{E+m+\beta \mathbf{\alpha P}}{\left( 2E\left( E+m\right) \right) ^{1/2}}%
.$ For the Dirac particles we have two energy subspaces $\varepsilon _{\pm }$
of dimension 2 corresponding to the positive and negative energy. Now with
the correspondence $\mathbf{Q}\rightarrow \mathbf{R},$ $\mathbf{K}%
\rightarrow \mathbf{p}$ and formula Eq. $\left( \ref{epsilonh2}\right) ,$
one easily sees that the diagonal matrix can be written%
\begin{equation}
\varepsilon \left( \mathbf{p,r}\right) =\beta E\left( \mathbf{p}\right)
+V\left( \mathbf{r}\right) +\beta \Phi  \label{EDirac}
\end{equation}%
The position operator is given by the ($4\times 4$) matrix $\mathbf{r=R+}%
\frac{\hbar \mathbf{p\times \Sigma }}{2E\left( E+m\right) }$ with $\mathbf{%
\Sigma =1\otimes \sigma }$ where $\mathbf{\sigma }$ are the Pauli matrices.
The band index of the scalar potential has been transferred to the matrix $%
\beta $, and we have $G^{ij}=\frac{1}{4E^{2}}g^{ij}$ and $T^{ij}=\frac{1}{%
8E^{3}}g^{ij}$ with the notation $g^{ij}=\delta ^{ij}-\frac{p^{i}p^{j}}{E^{2}%
},$ so that finally we can write 
\begin{equation}
\Phi =\frac{\hbar ^{2}}{8E^{2}}g^{ij}(\mathbf{\partial }_{i}\mathbf{\partial 
}_{j}V\mathbf{+}\frac{1}{E}\mathbf{\partial }_{i}V\partial _{j}V)
\label{phidirac}
\end{equation}%
If for central potential one can neglect the contribution $\frac{1}{E}%
\mathbf{\partial }_{i}V\partial _{j}V$, this is not always true and for some
potentials both terms in Eq.\ $\left( \ref{phidirac}\right) $ can be of the
same magnitude. In fact for constant electric field $V=-e\mathcal{E}\mathbf{%
.R}$, the first term vanishes and $\Phi =\frac{e^{2}\hbar ^{2}}{8E^{2}}g^{ij}%
\mathcal{E}_{i}\mathcal{E}_{j}$.

In the non-relativistic limit $\mathbf{p}<<m$, $\Phi $ becomes $\Phi \approx 
\frac{\hbar ^{2}}{8m^{2}}\left( \mathbf{\Delta }V\mathbf{+}\frac{1}{m}(%
\mathbf{\nabla }V)^{2})\right) +O(\hbar ^{2}p^{2}/m^{4})$ which gives two
contributions. The first one is the usual Darwin term $\frac{\hbar ^{2}}{%
8m^{2}}\mathbf{\Delta }V$ traditionally obtained as the result of the Foldy
Wouthuysen transformation expanded in power of $1/m.$ The second term $\frac{%
\hbar ^{2}}{8m^{3}}\left( \mathbf{\nabla }V\right) ^{2}$ of higher order in $%
1/m$ is usually not considered in the Foldy Wouthuysen approach. It is also
interesting to note that the external potential in the non relativistic
limit can be expanded as $V\left( \mathbf{r}\right) \approx V\left( \mathbf{R%
}\right) +\frac{\hbar }{4m^{2}}\mathbf{\Sigma \mathbf{.}}\left( \mathbf{%
\nabla }V\times \mathbf{p}\right) +O(\hbar ^{2}p^{2}/m^{4})$ where $\frac{%
\hbar }{4m^{2}}\mathbf{\Sigma \mathbf{.}}\left( \mathbf{\nabla }V\times 
\mathbf{p}\right) $ is the spin-orbit coupling term. Therefore the
Hamiltonian can be approximated as

\begin{eqnarray}
\varepsilon &\approx &\beta \left( m+\frac{\mathbf{P}^{2}}{2m}-\frac{\mathbf{%
P}^{4}}{8m^{3}}\right) +V\left( \mathbf{R}\right) +\frac{\hbar }{4m^{2}}%
\mathbf{\Sigma \mathbf{.}}\left( \mathbf{\nabla }V\times \mathbf{p}\right) 
\notag \\
&&\mathbf{+}\frac{\hbar ^{2}}{8m^{2}}\beta \left( \mathbf{\Delta }V\mathbf{+}%
\frac{1}{m}(\mathbf{\nabla }V)^{2}\right)  \label{Hdirac}
\end{eqnarray}%
A Born-Oppenheimer treatment of the Dirac equation where the spin is the
fast variable and the momentum the slow one has led to the same Hamiltonian
Eq.\ $\left( \ref{Hdirac}\right) $ but without the scalar potential \cite%
{MATHUR}. This corresponds to the semiclassical approximation. The
additional electric-type potential $\Phi $ is a consequence of transitions
between energy levels. This is in agreement with the usual interpretation of
the physical origin of the Darwin term, the zitterbewegung phenomenon,
whereby the electron does not move smoothly but instead undergoes extremely
rapid small-scale fluctuations due to an interference between positive and
negative energy states.

\textit{Acknowledgement}. We are grateful to Prof. M. V. Berry for having
drawn our attention to this subject.

\end{document}